\documentclass[a4paper,11pt]{article}
\pdfoutput=1 

\usepackage{jinstpub} 
\usepackage{subcaption}
\usepackage[T1]{fontenc} 

\title{\bf{Neutrino oscillation parameter determination at INO-ICAL using track and hit information from GEANT}}

\author[a,b,1]{Jaydeep Datta, \note{Université libre de Bruxelles, Av. Franklin Roosevelt 50, 1050 Bruxelles, Belgium (Present address)}}
\author[c,2]{Bana Singh, \note{Indian Institute of Science Education and Research,  Mohali, Punjab 140306, India (Present address)}}
\author[c]{S. Uma Sankar}

\affiliation[a]{Saha Institute of Nuclear Physics,\\Bidhannagar, Kolkata 700064, India}
\affiliation[b]{Homi Bhabha National Institute,\\Anushakti Nagar, Mumbai 400094, India}
\affiliation[c]{Department of Physics, Indian Institute of Technology Bombay,\\Mumbai 400076, India}
\emailAdd{uma@phy.iitb.ac.in}

\abstract{We study the capability of INO-ICAL to determine the neutrino oscillation parameters $|\Delta m^2_{31}|$ and $\sin^2 \theta_{23}$.
We do not use any generator level information.
Instead, we process the generated atmospheric neutrino events through GEANT4 simulation of the detector and the event reconstruction framework.
Among the outputs of this framework, only the momentum and direction of the longest track were used in a previous study by other authors.
In this study, in addition to these two variables, we consider a third variable based on additional hits, which arise due to hadrons in the event.
We show that the inclusion of this variable leads to a $30\%$ reduction in the uncertainty of $|\Delta m^2_{31}|$ for a 5-year run of ICAL.
We find that doubling the exposure time leads to an additional $30\%$ reduction in the uncertainties of both $|\Delta m^2_{31}|$ and $\sin^2 \theta_{23}$.}

\begin{document} 
\maketitle
\flushbottom

\section{Introduction}
\label{sec:intro}

The pioneering water Cerenkov detectors, IMB \cite{Casper:1990ac, BeckerSzendy:1992hq} and Kamiokande \cite{Hirata:1992ku,Fukuda:1994mc}, observed a deficit in the up-going muon events generated by atmospheric neutrinos relative to down-going events.
It was hypothesized that the deficit was caused by neutrino oscillations.
This hypothesis was confirmed by the zenith angle dependence of event rates observed by Super Kamiokande \cite{Fukuda:1998mi}.
Analysis of Super Kamiokande \cite{Super-Kamiokande:2017yvm} data gave 
\begin{equation}
    \Delta m^2_{\rm atm} \approx 3 \times 10^{-3}~ {\rm eV}^2 ~~~{\rm and} ~~~ \sin^2 2\theta_{\rm atm} \approx 1, 
\end{equation}
for atmospheric neutrino oscillation parameters.
A deficit in solar neutrinos, observed earlier, was also explained in terms of neutrino oscillations with 
\begin{equation}
    \Delta m^2_{\rm sol} \approx (2 - 20) \times 10^{-5}~ {\rm eV}^2 ~~~{\rm and} ~~~ \sin^2 \theta_{\rm sol} \approx 0.3, 
\end{equation}
as the solar neutrino oscillation parameters \cite{Bahcall:2004ut}.

It is possible to account for both solar and atmospheric neutrino oscillations in a three flavor oscillation framework.
The three neutrino flavor states, $\nu_{\alpha}$ ($\alpha = e, \mu, \tau$), mix to form three mass eigenstates $\nu_i$ ($i=1,2,3$)
with masses $m_i$. There are two independent mass squared differences.
Without loss of generality, they can be defined as $\Delta m^2_{21} = m^2_2 - m_1^2 \equiv \Delta m^2_{\rm sol}$ and 
$\Delta m^2_{31} = m^2_3 - m_1^2 \equiv \Delta m^2_{\rm atm}$.
Since $\Delta m^2_{\rm sol} \ll \Delta m^2_{\rm atm}$, the two larger mass square differences are nearly equal
$\Delta m^2_{32} \approx \Delta m^2_{31}$.
The unitary mixing matrix (PMNS matrix), connecting $\nu_{\alpha}$ to $\nu_i$, is parametrized in terms of three mixing angles ($\theta_{12}, \theta_{13}$ and $\theta_{23}$) and one CP violating phase $\delta_{\rm CP}$.
The CHOOZ experiment \cite{CHOOZ:1997cow, CHOOZ:1999hei} set the upper limit $\sin^2 2\theta_{13} \leq 0.1$ for values of $\Delta m^2_{31}$ > $10^{-3}$ eV$^2$.
Given this constraint, it can be shown that $\theta_{12} \approx \theta_{\rm sol}$  and $\theta_{23} \approx \theta_{\rm atm}$ \cite{Narayan:1997mk}.
Later experiments with man-made sources have confirmed the three flavor oscillation picture and have performed precision measurements of the mass squared differences and the mixing angles \cite{Esteban:2020cvm}.
Solar neutrino data shows that $\nu_e$ has a larger overlap with the lighter mass eigenstate, implying that $\Delta m^2_{21} > 0$ and $\theta_{12} < 45^\circ$ without loss of generality \cite{Fogli:2005cq,Nizam:2019eim}.
On the other hand, the sign of $\Delta m^2_{31}$ is one of the three unknowns of the three flavor neutrino oscillation paradigm 
\cite{ICAL:2015stm}.
These are
\begin{itemize}
    \item the CP violating phase $\delta_{\rm CP}$,
    \item the sign of $\Delta m^2_{31}$ (mass hierarchy) and
    \item the octant of $\theta_{23}$.
\end{itemize}
Atmospheric and accelerator neutrino data measure $\sin^2 2\theta_{23}$ to be very close to one \cite{MINOS:2014rjg, T2K:2014ghj, NOvA:2019cyt} which gives rise to the last problem.
A number of experiments are either underway or proposed to determine these unknowns.
Among them, the Iron Calorimeter (ICAL) at India-based Neutrino Observatory (INO) \cite{ICAL:2015stm} has a unique ability to determine the mass hierarchy \cite{Petcov:2005rv, Gandhi:2007td, Ghosh:2012px, Ajmi:2015uda}.

ICAL, a magnetized iron calorimeter, consists of iron slabs of thickness 40 mm and interspersed with resistive plate chambers.
A muon neutrino (anti-neutrino) interacting in the iron slab through charged current (CC) interaction creates a negative (positive) muon, which leaves a track in the detector.
The magnetic field of the detector bends these tracks, which leads to an identification of the charge of the leptons.
This ability to measure the neutrino and anti-neutrino event rates separately is crucial in mass hierarchy determination.
In addition, some other physics goals of ICAL are
\begin{itemize}
    \item observing the minimum in the $(L/E)$ distribution of the muon event rate \cite{Kumar:2020wgz},
    \item measuring the density profile of earth \cite{Kumar:2021faw},
    \item searching for CPT violation \cite{Datta:2003dg} and
    \item searching for sterile neutrinos \cite{Behera:2016kwr}.
\end{itemize}
In addition, ICAL can also improve the precision of atmospheric neutrino oscillation parameters. 

The first studies \cite{Thakore:2013xqa} of oscillation parameter precision at ICAL were done  with event samples binned in muon energy and muon direction, which is characterized by the cosine of the nadir angle of the muon.
The muon kinematic information used in these studies was taken from the event generator and smeared with resolution functions. 
Later studies \cite{Devi:2014yaa} binned in an additional variable $E_{\rm had} = E_{\nu}-E_{\mu}$ and showed that the precision improves significantly.
The kinematical variables used in these studies will not be available in a real experiment.
The realistic kinematical variables are the reconstructed track energy and track direction, which are obtained by a full GEANT4 simulation \cite{GEANT4:2002zbu} of atmospheric neutrino events in ICAL and processing the simulation output through the event reconstruction code \cite{Bhattacharya:2014tha}.
In an earlier work, Rebin et al. \cite{Rebin:2018fdl} studied the oscillation parameter precision using the event distributions binned in these two track variables for a simulated 5-year event sample as data.
For theoretical prediction, they simulated a 1000-year event sample and scaled it down by 200, thus taking into account the statistical fluctuations present in the data.
Their results are summarized in table \ref{Chap10:table1}.
\begin{table}[ht]
	\centering
	\begin{tabular}{@{}|c|c|c|c|c|@{}}
		\hline
		Parameter & Input Value & Reconstructed & 2~$\sigma$ Range & 3~$\sigma$ Range \\
		&& Best Fit point & & \\
		\hline
		$|\Delta m^2_{31}|$(*10$^{-3}$ eV$^2$)&2.32&2.32 & 2.03 (1.68 - 3.71) & 4.07 (1.40 - 5.47)\\
		\hline
		$\sin^2\theta_{23}$& 0.5 & 0.496 & 0.38 (0.34 - 0.72) & 0.48 (0.29 - 0.77)\\
		\hline
	\end{tabular}
	\caption{\label{Chap10:table1} Results of 2-variable analysis taken from figure 7 of ref. \cite{Rebin:2018fdl}. The selected event sample is for a 5-year exposure, and no event selection cuts were imposed.}
	\end{table}

In this work, we analyzed the simulated atmospheric neutrino data by binning it in three variables: 
magnitude of track momentum, track direction and a third variable which is a measure of hadronic energy in the event
and did a comparison of this analysis with the analysis based only on
track momentum and track direction. 
We also studied how the parameter precision improves with increased exposure.
\section{\label{Chap10:sec1}Methodology}
\subsection{\label{Chap10:sec1a}Data and theory samples}
The analysis started with 500 years of un-oscillated events generated by NUANCE \cite{Casper:2002sd}, with Kamiokande fluxes as input \cite{Honda:2011nf}.
From this, a 5-year event sample was isolated.
The survival probabilities, $P_{\mu \mu}$ and $P_{\bar{\mu} \bar{\mu}}$, were calculated using the three flavor oscillation code \textit{nuCraft} \cite{Wallraff:2014qka}.
The following values of neutrino oscillation parameters were used as inputs: $\sin^2 \theta_{12} = 0.310$, $\sin^2 \theta_{13} = 0.02240$, $\sin^2 \theta_{23} = 0.5$, $\Delta m^2_{31} = 2.32 \times 10^{-3}$ eV$^2$, $\Delta m^2_{21} = 7.39 \times 10^{-5}$ eV$^2$ and $\delta_{ CP}= 0$.
The accept/reject method was applied to $\nu_{\mu}(\bar{\nu}_{\mu})$ CC events using $P_{\mu \mu} (P_{\bar{\mu} \bar{\mu}})$ to obtain the muon events due to the survival of $\nu_{\mu}/\bar{\nu}_{\mu}$.
Because $\sin^2 \theta_{13} \ll 1$, the muon CC events occurring due to the oscillation of 
$\nu_{e}\to \nu_{\mu}(\bar{\nu}_{e} \to \bar{\nu}_{\mu})$ formed a small fraction (about <1\%) of the total number.
Hence, they were not included in this analysis. 
This 5-year event sample was labelled \textit{data}.

To calculate the corresponding theoretical event sample, with test values of neutrino oscillation parameters, the following method was followed.
A test value of $|\Delta m^2_{31}|$ from the range $(1,5) \times 10^{-3}$ eV$^2$ and a test value of the $\sin^2 \theta_{23}$ in the range $(0,1)$ were chosen.
The test values of $|\Delta m^2_{31}|$ were varied systematically in steps of $0.2 \times 10^{-3}$ eV$^2$ in the range $(1,5) \times 10^{-3}$ eV$^2$ and $\sin^2 \theta_{23}$ in steps of $0.02$ in the range $(0,1)$.
The other neutrino oscillation parameters were kept fixed at the input values, which were used to generate the data event sample.
The remaining 495 year un-oscillated muon event sample was converted into an oscillated event sample through accept/reject method using survival probabilities calculated with \textit{nuCraft} \cite{Wallraff:2014qka}.
This sample was scaled by 99 to obtain a test event sample corresponding to five years, which was labelled as \textit{theory}.

\subsection{\label{Chap10:sec1b}Variables for analysis}

Muons, being minimum ionizing particles, pass through many layers of iron, leaving behind localized hits in the Resistive Plate Chambers (RPCs).
Using this hit information, the track of the muon can be reconstructed.
Because of the magnetic field, this track will be curved and the bending of the track is opposite for negative and positive muons.
Thus, ICAL can distinguish between the CC interactions of $\nu_{\mu}$ and $\bar{\nu}_{\mu}$.
If a track is reconstructed, the event is considered to be a CC interaction of $\nu_{\mu} /\bar{\nu}_{\mu}$.
The charge, the momentum and the initial direction ($\cos \theta_{ track}$) of a reconstructed track are also 
calculated from the track properties \cite{Bhattacharya:2014tha}.
Electrons, positrons and photons lose their energy very quickly and leave no track.
Pions pass through a few layers, and it is difficult to construct a track for them unless they are of very high energy.
Baryons have very low kinetic energy and can pass through one or two layers.
Therefore, the hadrons produced in these reactions leave a few $(0 - 10)$ scattered hits \cite{Devi:2013wxa,Nizam:2019jdi}.
Thus, the GEANT4 simulation of a typical $\nu_{\mu}/\bar{\nu}_{\mu}$ CC event consists of a track and a few hadron hits.

Events for which one or more tracks were reconstructed were stored along with the charge, the momentum and the 
initial direction of the track with the largest momentum.
For about 90\% of the events, only one track was reconstructed.
In the other 10\%, two or more tracks were reconstructed.
The leftover hits, that is the hits which were not used for track reconstruction, were labelled hadron hits.
In dealing with the hadron hits, the problem of ghost hits had to be considered.
If the actual hits occurred at $(x_1, y_1)$, $(x_2,y_2)$ and $(x_3,y_1)$ the simulation only told us that the $x$-strips $x_1, x_2, x_3$ and the $y$-strips $y_1, y_2$ are hit.
This led to the possibility of there being six hits, three of which were ghost hits.
This problem was solved very simply by defining the number of hadron hits to be the maximum of 
\textit{(number of $x$-strips, number of $y$-strips)} with hits. 

CPT conservation implies that the values of the mass squared differences and mixing angles of neutrinos and anti-neutrinos are the same.
Therefore, we did our initial analysis without taking muon charge into account.
We considered $\nu_{\mu}$ and $\bar{\nu}_{\mu}$ CC events together, which also led to an improvement in the statistics.
Since muon charge information was available, we performed a second iteration of our analysis where we considered $\nu_{\mu}$ and $\bar{\nu}_{\mu}$ CC events separately.
The results of both analyses were very similar.
The time information of the events was not used explicitly.
The timestamps of hits, however, were used by the reconstruction program in determining the track direction.

\subsection{\label{Chap10:sec1c}Analysis methods}

The analysis was done using two different binning schemes.
These schemes are labelled as \textit{2-variable analysis} and \textit{3-variable analysis}.
\begin{itemize}
	\item \underline{2-variable analysis}: In this analysis method, the events were binned according to the magnitude of the reconstructed track momentum and the reconstructed track direction.
	For the events where more than one track was reconstructed, the track momentum and direction were those of the longest track.
	There were 10 momentum bins given by $(0.1,0.5)$, $(0.5,1.0)$, $(1.0,1.5)$, $(1.5,2.0)$, $(2.0,2.5)$, $(2.5,3.0)$, $(3.0,5.0)$, $(5.0,10.0)$, $(10.0,20.0)$, $(20.0,100.0)$ GeV/$c$.
	At higher energies, the number of atmospheric neutrino events becomes quite small.
	To have a sufficient number of events in each bin, larger bin sizes were chosen at higher energies.
	We tried a number of plausible binning schemes and selected the one which gave the best results. 
	From the allowed range $(-1, 1)$ of the cosine of the nadir angle of the track momentum, 40 direction bins were chosen in steps of $0.05$.
	This analysis used 400 bins in the double differential distribution of the two variables.
	\item \underline{3-variable analysis}:
	The three variables used in this analysis are, $E$, $\cos\theta_{\rm track}$ and the number of non-track hits.
	Here, $E$ was the magnitude of the track momentum for single track events, and it was the sum of the magnitudes of track momenta for multi-track events.
	The variable $\cos\theta_{\rm track}$ was the direction of the reconstructed track for single track events, and it was the direction of the longest track for multi-track events.
	We classified events into four classes based on $E$ and $|\cos \theta_{\rm track}|$. 
	They were:
	\begin{enumerate}
	\item High energy, horizontal events with $E~>$ 10 GeV and $|\cos \theta_{\rm track}|~<$ 0.3, with 2 bins in $E$ and 12 bins in $|\cos \theta_{\rm track}|$ for a total of 24 bins.
	\item High energy, vertical events with $E~>$ 10 GeV and $|\cos \theta_{\rm track}|~>$ 0.3, with 2 bins in $E$ and 28 bins in $|\cos \theta_{\rm track}|$ for a total of 56 bins.
	\item Low energy, horizontal events with $E~<$ 10 GeV and $|\cos \theta_{\rm track}|~<$ 0.3, with 8 bins in $E$ and 12 bins in $|\cos \theta_{\rm track}|$ for a total of 96 bins.
	\item Low energy, vertical events with $E~<$ 10 GeV and $|\cos \theta_{\rm track}|~>$ 0.3, with 8 bins in $E$ and 28 bins in $|\cos \theta_{\rm track}|$ for a total of 224 bins.
	\end{enumerate}
\end{itemize}
For atmospheric neutrinos, the number of events at high energies is small and for ICAL the reconstruction ability for horizontal events is poor.
The number of events in each bin of the first three classes were too small to be further subdivided based on non-track hits.
Hence, the events in these 176 bins were binned only in the two variables $E$ and $\cos \theta_{\rm track}$.
Binning in the third variable, non-track hits, was done only for the 224 bins of low energy, vertical events.
The bins used for the third variable are $(0,4)$, $(5,10)$ and $> 10$. 

The \textit{data} and the \textit{theory} samples were binned according to the scheme described above.
Each entry of the \textit{theory} sample was divided by 99.
As the theory samples were obtained by down scaling a very large generated sample, the effect of fluctuations was negligible.
On the other hand, event number fluctuations were expected to distort the event distributions in each of the kinematic variables in the \textit{data} sample.

The survival probabilities depend on the distance travelled by the neutrinos.
An accurate estimate of this distance requires a precise value of $\cos \theta_{\rm track}$.
A bin size of 0.2 was chosen for this variable in ref. \cite{Rebin:2018fdl}.
On the other hand, we chose a much finer bin size of 0.05,
expecting this to lead to a much better precision in $|\Delta m^2_{31}|$.

\subsection{\label{Chap10:sec1d}Systematic errors and \texorpdfstring{$\chi^2$}{TEXT}%
 ~ calculation}

The \textit{theory} event sample was calculated with Kamiokande atmospheric neutrino fluxes and cross-sections 
implemented in the NUANCE event generator as inputs.
However, these inputs had uncertainties associated with them.
They were taken into account as systematic errors through the method of pulls \cite{Gonzalez-Garcia:2004pka}.
Usually three systematic errors were considered: a) normalization error, b) energy-dependent tilt error and c) zenith angle dependent error.
The second and third errors were defined in terms of neutrino energy and direction, respectively.
But in our analysis, the variables used are track momentum and direction.
Hence, the systematic errors specified as functions of neutrino energy and direction had to be transformed 
into functions of track momentum and track direction.
In ref. \cite{Nizam:2019eim} transfer matrices were constructed to convert the neutrino energy-dependent tilt error 
to track momentum-dependent tilt error and neutrino direction-dependent error to reconstructed track direction-dependent error.
These two errors were assumed to be independent of each other.
As described in section \ref{Chap10:sec1c}, some events were binned using the non-track hit information. 
An overall $0.05$ systematic error was considered for each of those bins.
This was assumed to be independent of the momentum and direction of the longest track.

The number of events in bin $ijk$, $N_{ijk}^{\rm test}$ was defined as
\begin{equation}
N_{ijk}^{\rm test} = N_{ijk}^{\rm theory} [1 + \pi_{ijk}^{l} \xi_{l}],
\end{equation}
where $\xi_l$ was the pull variable for the systematic error $l$ and it was varied in the range $-3$ to $3$ in steps of 0.5.
Each of the pull coefficients $\pi_{ijk}^{l}$ was assumed to be small compared to 1.
The first three systematic errors were flux normalization dependent error ($\pi_{\rm norm}$), track momentum dependent 
tilt error ($\pi_{\rm tilt}$), track direction dependent error ($\pi_{\rm dir}$) and the last one was systematic error 
($\pi_{\rm had}$) taken for each non-track hit information bin in case of 3-variable analysis.
A constant value of 0.2 was assumed for $\pi_{\rm norm}$ \cite{Gonzalez-Garcia:2004pka}.
Following \cite{ParticleDataGroup:2020ssz}, the $\chi^2$ between the data and test event samples was
\begin{equation}
\chi^2 = \sum_{ijk} \left[2 \left\{ \left(N_{ijk}^{\rm test} - N_{ijk}^{\rm data} \right) - N_{ijk}^{\rm data} \ln\left(\frac{N_{ijk}^{\rm test}}{N_{ijk}^{\rm data}}\right) \right\} \right] +\xi_l^2.
\end{equation}
This $\chi^2$ was a function of the test values of $|\Delta m^2_{31}|$ and $\sin^2 \theta_{23}$.
For each pair of test values of $|\Delta m^2_{31}|$ and $\sin^2 \theta_{23}$, $\chi^2$ was computed, and the minimum 
$\chi^2$ was evaluated.
The inputs to the NUANCE event generator were best fit results of some previous experiments.
Large values of $\xi_l$ pulled these inputs away from these best fit values.
The priors $\xi_l^2$ prevented the pull variables from taking too large values.
The test values of $|\Delta m^2_{31}|$ and $\sin^2 \theta_{23}$ corresponding to minimum $\chi^2$ were their best fit values.

\subsection{\label{Chap10:sec1e}Procedure to calculate best fit point, \texorpdfstring{2$\sigma$}{2 s} and \texorpdfstring{3$\sigma$}{3 s} range}

The procedure described above was carried out for 25 mutually independent 5-year muon data samples, which were 
extracted randomly from the 500-year sample.
The value of $\chi^2$ for each of the data samples was plotted as a function of test $|\Delta m^2_{31}|$ and 
as a function of test $\sin^2 \theta_{23}$.
From these plots, the values of $|\Delta m^2_{31}|$ and $\sin^2 \theta_{23}$ which minimize $\chi^2$ were 
obtained for each data sample.
These values of the oscillation parameters were noted down as the best fit point for that data sample.
The values of $|\Delta m^2_{31}|$ and $\sin^2 \theta_{23}$ for which $\Delta \chi^2 = \chi^2 - \chi^2_{\rm min}$ is 4 (9) 
defined the 2$\sigma$ (3$\sigma$) range of the oscillation parameters.
We chose not to define 1$\sigma$ range of the oscillation parameters using a similar definition because 
such ranges show a wide variation due to the fluctuations in the data sample.
These ranges were calculated for each data sample, along with the best fit point.
It was observed that for some samples, the upper bound of 2$\sigma$ and 3$\sigma$ for $|\Delta m^2_{31}|$ lay
outside the corresponding test value range.
For those samples it was assumed that the upper bound of 2$\sigma$ was $7\times10^{-3}$ eV$^2$ and that for 
3$\sigma$ range was $10\times10^{-3}$ eV$^2$.
We computed the averages of the best fit parameter values, 2$\sigma$ ranges and 3$\sigma$ ranges over the 25 sets, 
for the purpose of comparing the results of the two analysis methods. 
In the following subsections, the results for 5-year and 10-year exposure time and 2-variable analysis and 3-variable 
analysis are discussed.

\section{\label{Chap10:sec2}Results}
\subsection{\label{Chap10:sec2b}Comparison of the 2-variable and the 3-variable analysis methods}

\begin{figure}
    \centering
    \includegraphics[width=\linewidth]{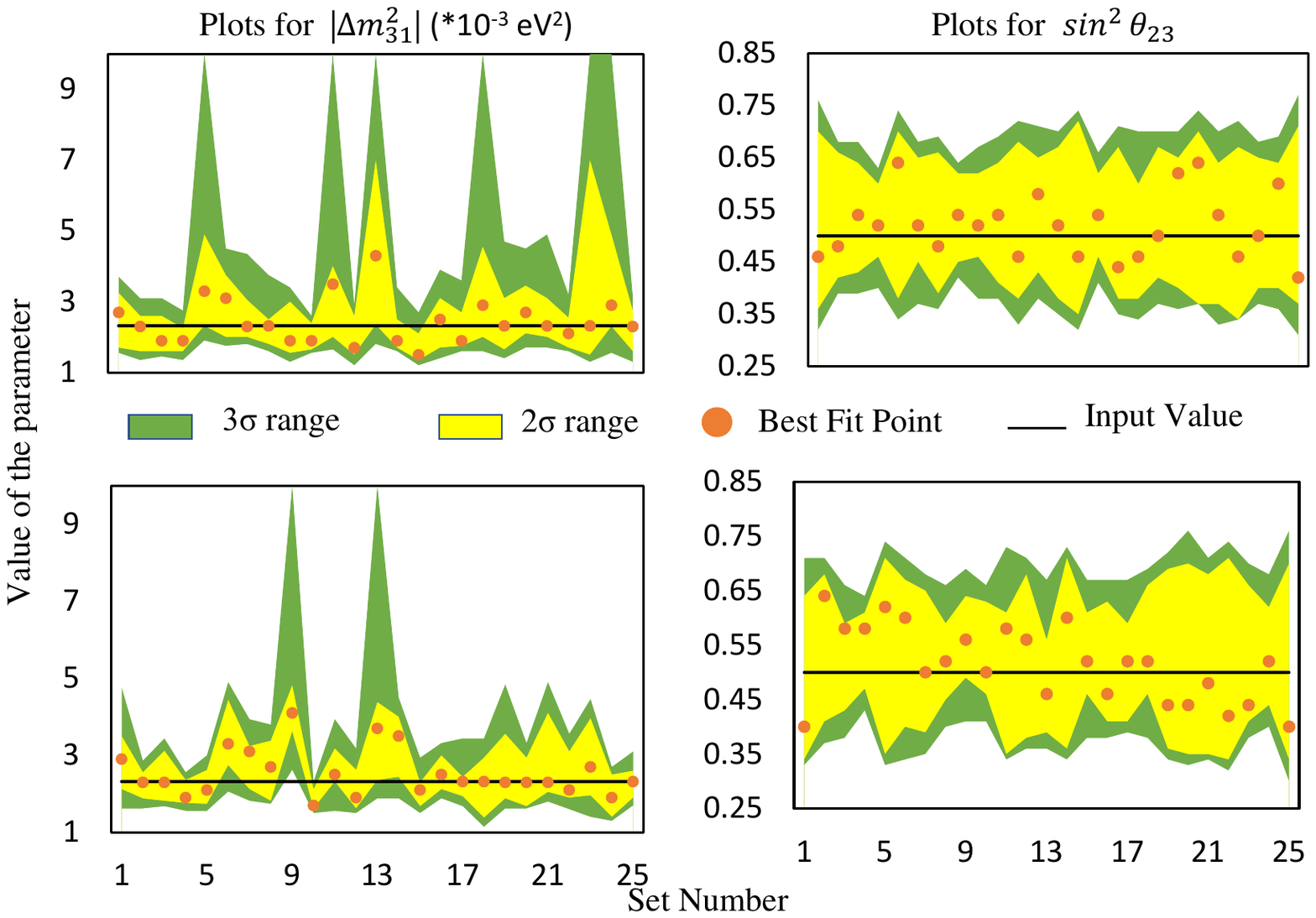}
    \caption{Best fit point, 2$\sigma$ upper and lower limits and 3$\sigma$ upper and lower limits of $|\Delta m^2_{31}|$ 
and $\sin^2 \theta_{23}$ for the 25 mutually exclusive data sets we considered. The exposure considered was five years. 
The top panel is for the 2-variable analysis method and the bottom-panel is for the 3-variable analysis method. 
The left panel is for $|\Delta m^2_{31}|$ and the right panel is for $\sin^2 \theta_{23}$. Of the 25 sets, 15 sets give 
smaller allowed range of $|\Delta m^2_{31}|$ for the 3-variable analysis compared to the 2-variable analysis, and 12 sets 
give smaller allowed range of $\sin^2 \theta_{23}$ for the 3-variable analysis compared to the 2-variable analysis.}
    \label{fig:1}
\end{figure}

\begin{figure}
    \centering
    \includegraphics[width=\linewidth]{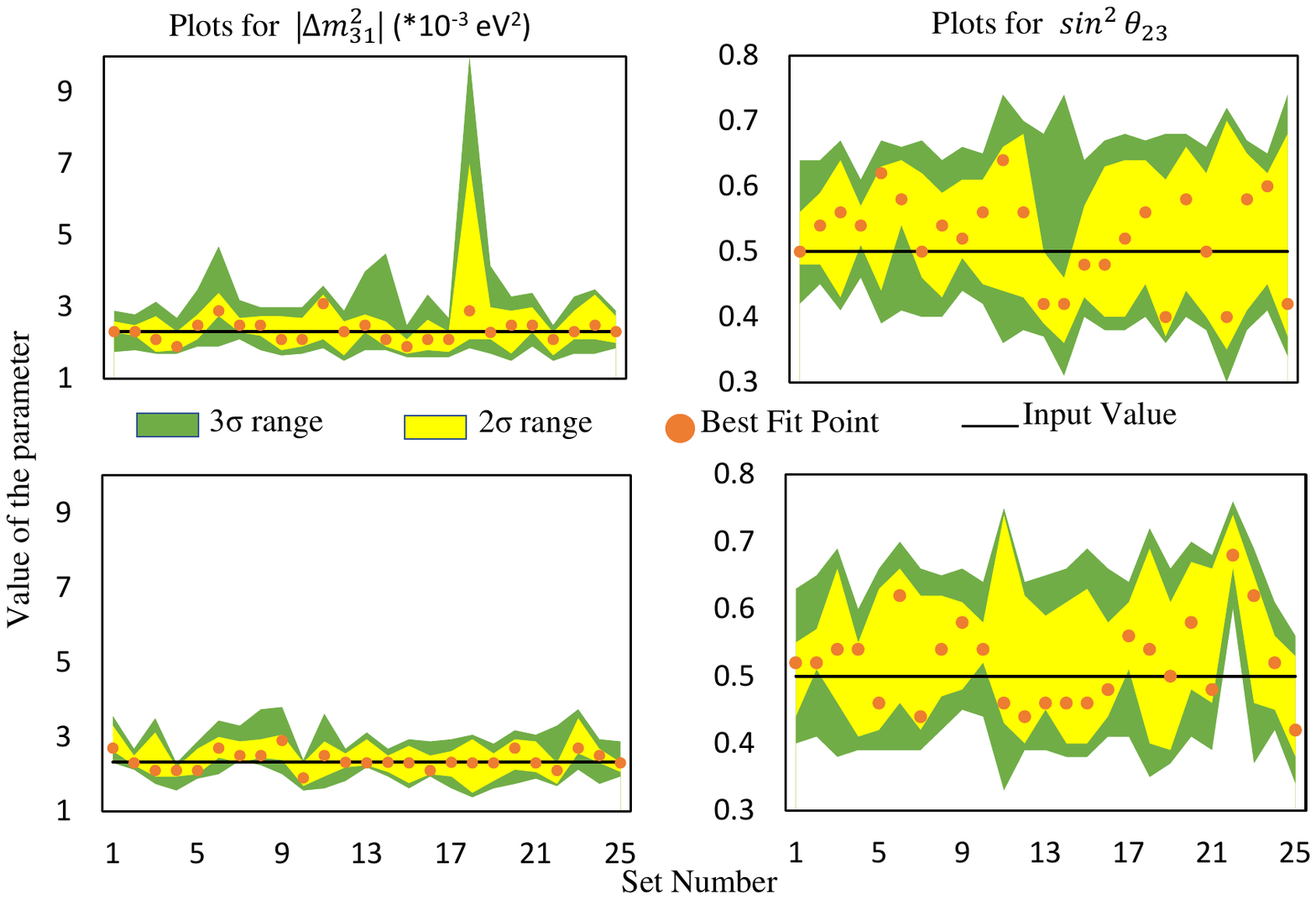}
    \caption{Best fit point, 2$\sigma$ upper and lower limits and 3$\sigma$ upper and lower limits of $|\Delta m^2_{31}|$ 
and $\sin^2 \theta_{23}$ for the 25 mutually exclusive data sets we considered. The exposure considered was 10 years. The 
top panel is for the 2-variable analysis method and the bottom-panel is for the 3-variable analysis method. The left panel 
is for $|\Delta m^2_{31}|$ and, the right panel is for $\sin^2 \theta_{23}$. Of the 25 sets, 18 sets give smaller allowed 
range of $|\Delta m^2_{31}|$ for the 3-variable analysis compared to the 2-variable analysis, and 13 sets give smaller 
allowed range of $\sin^2 \theta_{23}$ for the 3-variable analysis compared to the 2-variable analysis.}
    \label{fig:2}
\end{figure}

The results for 5-year exposure are summarized in tables \ref{Chap10:table2}.
\begin{table}
	\centering
	\begin{tabular}{|c|c|c|c|c|c|c|}
		\hline
		\multicolumn{7}{|c|}{Average of 25 Sets}\\
		\hline
		Parameter  & Binning  & $3\sigma$ & $2\sigma$& Best & $2\sigma$  & $3\sigma$\\
		& Scheme & Lower & Lower & Fit & Upper & Upper\\
		& & Bound & Bound & Point & Bound & Bound\\
		\hline
		$|\Delta m^2_{31}|$ & 2-variable & 1.53 & 1.80 & 2.43 & 3.43 & 5.13 \\
		\cline{2-7}
		$(*10^{-3} eV^2)$ & 3-variable & 1.67 & 1.99 & 2.53 & 3.19 & 4.13\\
		\hline
		$\sin^2 \theta_{23}$ & 2-variable & 0.36 & 0.40 & 0.52 & 0.66 & 0.70\\
		\cline{2-7}
		& 3-variable & 0.36 & 0.40 & 0.51 & 0.65 & 0.70\\
		\hline
	\end{tabular}
	\caption{\label{Chap10:table2}Comparison of average best fit point, 2$\sigma$ and 3$\sigma$ allowed ranges of 
$|\Delta m^2_{31}|$ and $\sin^2 \theta_{23}$ for the 25 mutually exclusive data sets obtained with the 2-variable and 
the 3-variable analysis methods. The average values were calculated for 5-year exposure time.}
\end{table}
The obtained results for $\sin^2 \theta_{23}$ and its allowed ranges in 2$\sigma$ and 3$\sigma$ are comparable for the 2-variable analysis and for the 3-variable analysis.
On the other hand, the allowed ranges for $|\Delta m^2_{31}|$ show a noticeable improvement for the 3-variable analysis compared to the 2-variable analysis.
For example, the 2-variable analysis does not give an upper limit in the explored range at 2$\sigma$ in two of the 25 cases, whereas the 3-variable analysis does so for every case.
A comparison between the best fit points and the 2$\sigma$ and the 3$\sigma$ ranges of the parameters for the 2-variable analysis and the 3-variable analysis methods is shown in the figure \ref{fig:1}.
If the exposure is increased to ten years, the corresponding numbers are one and zero respectively as shown in figure \ref{fig:2}.
The results for ten-year exposure are summarized in table \ref{Chap10:table3}.
From table \ref{Chap10:table3}, we note that the average allowed ranges of $|\Delta m^2_{31}|$, both at 2$\sigma$ and at 3$\sigma$, are smaller for the 3-variable analysis compared to the 2-variable analysis.
\begin{table}
	\centering
	\begin{tabular}{|c|c|c|c|c|c|c|}
		\hline
		\multicolumn{7}{|c|}{Average of 25 Sets}\\
		\hline
		Parameter  & Binning  & $3\sigma$ & $2\sigma$& Best   & $2\sigma$  & $3\sigma$\\
		& Scheme & Lower & Lower & Fit & Upper & Upper\\
		& & Bound & Bound & Point & Bound & Bound\\
		\hline
		$|\Delta m^2_{31}|$ & 2-variable & 1.74 & 2.01 & 2.35 & 2.92 & 3.54\\
		\cline{2-7}
		$(*10^{-3} eV^2)$ & 3-variable & 1.86 & 2.09 & 2.37 & 2.77 & 3.08\\
		\hline
		$\sin^2 \theta_{23}$ & 2-variable & 0.39 & 0.43 & 0.52 & 0.62 & 0.67\\
		\cline{2-7}
		& 3-variable & 0.40 & 0.45 & 0.52 & 0.62 & 0.66\\
		\hline
	\end{tabular}
	\caption{\label{Chap10:table3}Comparison of average best fit point, 2$\sigma$ and 3$\sigma$ allowed ranges of 
$|\Delta m^2_{31}|$ and $\sin^2 \theta_{23}$ for the 25 mutually exclusive data sets obtained with the 2-variable and 
the 3-variable analysis methods. The average values were calculated for 10-year exposure time.}
\end{table}

\subsection{Analysis with independent \texorpdfstring{$\mu^-$}{TEXT} %
and \texorpdfstring{$\mu^+$}{TEXT} %
data sets}

\begin{figure}[h]
    \centering
    \includegraphics[width=\linewidth]{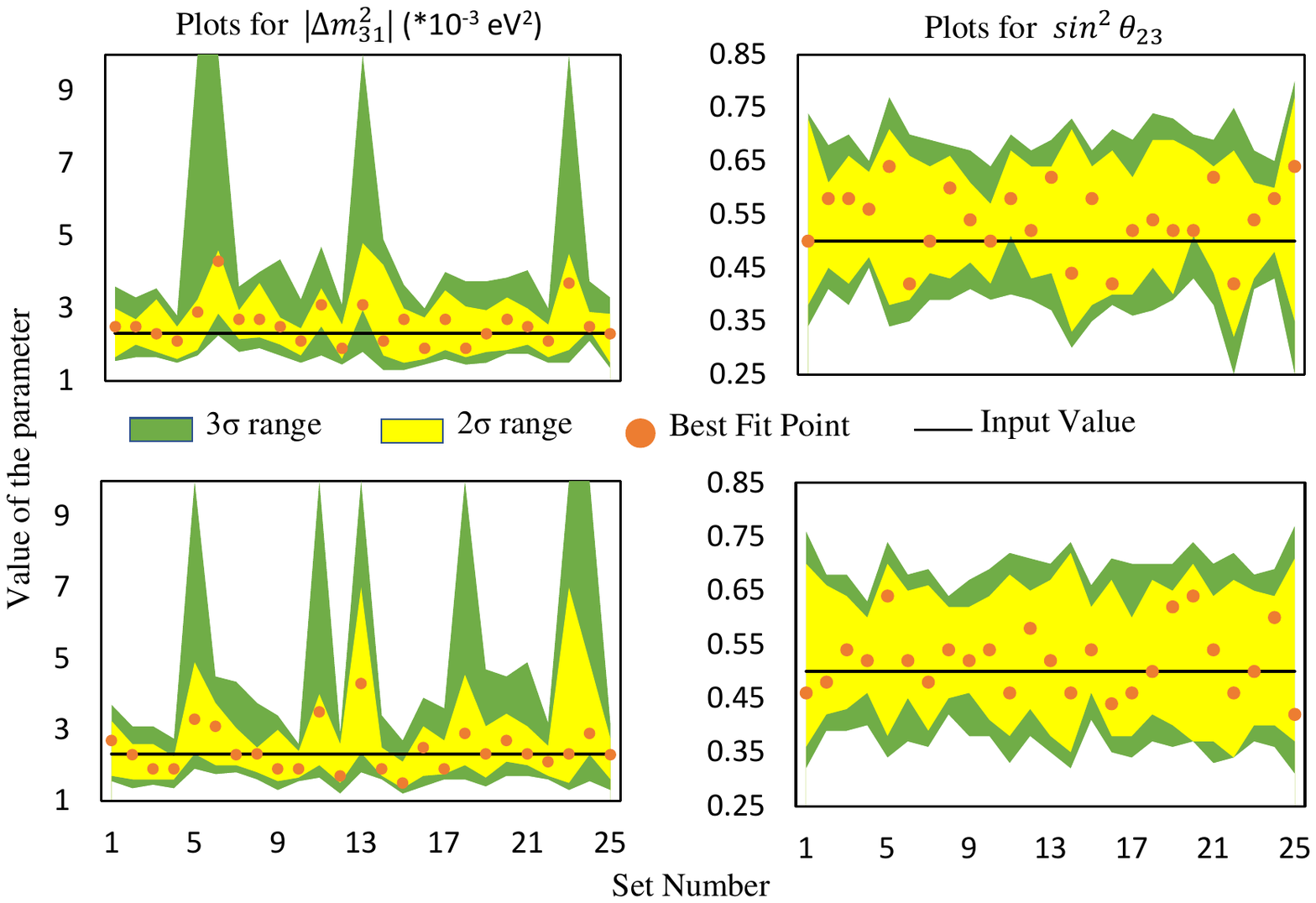}
    \caption{Best fit point, 2$\sigma$ upper and lower limits and 3$\sigma$ upper and lower limits of $|\Delta m^2_{31}|$ and $\sin^2 \theta_{23}$ for the 25 mutually exclusive data sets we considered. The exposure considered is five years. The top panel is for the 2-variable analysis with charge ID and the bottom-panel is for the 2-variable analysis without charge ID. The left panel is for $|\Delta m^2_{31}|$, and the right panel is for $\sin^2 \theta_{23}$. Of the 25 sets, 15 sets give smaller allowed range of $|\Delta m^2_{31}|$ for with charge ID compared to without charge ID analysis. The corresponding number for  $\sin^2 \theta_{23}$ is 15.}
    \label{fig:3}
\end{figure}

ICAL is a magnetized iron calorimeter, which is capable of determining the muon charge and hence we can study the oscillations of $\nu_{\mu}$ and $\bar{\nu}_{\mu}$ independently.
But the oscillation parameters are the same for neutrinos and anti-neutrinos as mentioned in section \ref{Chap10:sec1b}.
Therefore, the analysis described in the previous subsection binned the events using only track momentum and track direction, but not the charge ID of the track.
However, we wanted to explore if the inclusion of the charge ID of the track makes any difference to the obtained results.
To this end, we studied the determination of $|\Delta m^2_{31}|$ and $\sin^2 \theta_{23}$ by binning the events according to (a) track momentum, (b) track direction and (c) charge ID of the track.
This leads to 800 bins as opposed to 400 bins for the case where the charge ID was not considered. 
A $\chi^2$ analysis was carried out on event samples of 5-year exposure, using the procedure described in section \ref{Chap10:sec2}, for this binning where the charge ID is included.
The best fit point, 2$\sigma$ and 3$\sigma$ allowed range for the 25 sets are shown in figure \ref{fig:3} for analyses with and without charge ID.
The comparison between these two analysis methods are presented in table \ref{Chap10:table4}.
We note that the allowed ranges of parameters, without and with charge ID, are quite similar.
\begin{table}[h!]
	\centering
	\begin{tabular}{|c|c|c|c|c|c|c|}
		\hline
		\multicolumn{7}{|c|}{Average of 25 Sets}\\
		\hline
		Parameter & Binning  & $3~\sigma$ & $2~\sigma$& Best   & $2~\sigma$  & $3~\sigma$\\
		& Scheme & Lower & Lower & Fit & Upper & Upper\\
		& & Bound & Bound & Point & Bound & Bound\\
		\hline
		$|\Delta m^2_{31}|$ & Without charge ID & 1.53 & 1.80 & 2.43 & 3.43 & 5.13 \\
		\cline{2-7}
		$(*10^{-3} eV^2)$ & With charge ID & 1.63 & 1.93 & 2.56 & 3.22 & 4.69\\
		\hline
		$\sin^2 \theta_{23}$ & Without charge ID & 0.36 & 0.40 & 0.52 & 0.66 & 0.70\\
		\cline{2-7}
		& With charge ID & 0.37 & 0.42 & 0.54 & 0.66 & 0.70\\
		\hline
	\end{tabular}
	\caption{\label{Chap10:table4}Comparison of our results without and with charge ID. The results quoted are obtained using the 2-variable analysis method for a 5-year exposure time.}
\end{table}

In table \ref{Chap10:table6} we compare our results of 2-variable analysis with those of ref. \cite{Rebin:2018fdl}.
\begin{table}[ht]
	\centering
	\begin{tabular}{@{}|c|c|c|c|c|c|@{}}
		\hline
		Method & Parameter & Input & Reconstructed & 2$\sigma$  & 3$\sigma$\\
		&& Value & Best Fit point & Range & Range\\
		\hline
		Ref. \cite{Rebin:2018fdl} & $|\Delta m^2_{31}|$(*10$^{-3}$ eV$^2$)&2.32&2.32 & 2.03 (1.68 - 3.71) & 4.07 (1.40 - 5.47)\\
		\hline
		Ref. \cite{Rebin:2018fdl} & $\sin^2\theta_{23}$& 0.5 & 0.496 & 0.38 (0.34 - 0.72) & 0.48 (0.29 - 0.77)\\
		\hline
		5-yr, 2-var & $|\Delta m^2_{31}|$(*10$^{-3}$ eV$^2$)& 2.32 & 2.43 & 1.63 (1.80 - 3.43) & 3.60 (1.53 - 5.13) \\
		\hline
	    5-yr, 2-var & $\sin^2\theta_{23}$& 0.5 & 0.52 & 0.26 (0.40 - 0.66) & 0.34 (0.36 - 0.70)\\
		\hline
		\end{tabular}
	\caption{\label{Chap10:table6} Comparison of our results with those of ref. \cite{Rebin:2018fdl}.}
\end{table}
We note that the precision in $|\Delta m^2_{31}|$ of our analysis is better.
This improvement is expected because of the finer bin size in $\cos \theta_{\rm track}$, as mentioned in section \ref{Chap10:sec1c}.

\section{Conclusion}

\begin{table}[ht]
	\centering
	\begin{tabular}{@{}|c|c|c|c|c|c|@{}}
		\hline
		Method & Parameter & Input & Reconstructed & 2$\sigma$  & 3$\sigma$\\
		&& Value & Best Fit point & Range & Range\\
		\hline
		5-yr, 2-var & $|\Delta m^2_{31}|$(*10$^{-3}$ eV$^2$)& 2.32 & 2.43 & 1.63 (1.80 - 3.43) & 3.60 (1.53 - 5.13) \\
		\hline
	    5-yr, 2-var & $\sin^2\theta_{23}$& 0.5 & 0.52 & 0.26 (0.40 - 0.66) & 0.34 (0.36 - 0.70)\\
		\hline
		5-yr, 3-var & $|\Delta m^2_{31}|$(*10$^{-3}$ eV$^2$)& 2.32 & 2.53 & 1.20 (1.99 - 3.19) & 2.46 (1.67 - 4.13) \\
		\hline
	    5-yr, 3-var & $\sin^2\theta_{23}$& 0.5 & 0.51 & 0.25 (0.40 - 0.65) & 0.34 (0.36 - 0.70)\\
		\hline
		10-yr, 2-var & $|\Delta m^2_{31}|$(*10$^{-3}$ eV$^2$)& 2.32 & 2.35 & 0.91 (2.01 - 2.92) & 1.8 (1.74 - 3.54) \\
		\hline
	    10-yr, 2-var & $\sin^2\theta_{23}$& 0.5 & 0.52 & 0.19 (0.43 - 0.62) & 0.28 (0.39 - 0.67)\\
		\hline
		10-yr, 3-var & $|\Delta m^2_{31}|$(*10$^{-3}$ eV$^2$)& 2.32 & 2.37 & 0.68 (2.09 - 2.77) & 1.24 (1.86 - 3.10) \\
		\hline
	    10-yr, 3-var & $\sin^2\theta_{23}$& 0.5 & 0.51 & 0.17 (0.45 - 0.62) & 0.26 (0.40 - 0.66)\\
		\hline
	\end{tabular}
	\caption{\label{Chap10:table5} Comparison of uncertainty ranges from different methods.}
\end{table}
In this work, we estimated the ability of ICAL at INO to determine the neutrino oscillation parameters $|\Delta m^2_{31}|$ and $\sin^2 \theta_{23}$.
Most of the previous studies of this topic used generator level kinematic information, smeared by resolution functions.
In this work, we performed full GEANT4 simulation of atmospheric neutrino events.
Using the ICAL reconstruction code \cite{Bhattacharya:2014tha} on this simulated data, we extracted the kinematical variables of tracks and performed a $\chi^2$ analysis with the events binned in them.
In addition to performing a 2-variable analysis with binning only in track momentum and track direction, we did a 3-variable analysis where the number of non-track hits is introduced as a third variable.

From table \ref{Chap10:table5}, we see that both the 2$\sigma$ and 3$\sigma$ ranges of $|\Delta m^2_{31}|$ from the 3-variable analysis are smaller by a factor of 1.5 compared to the corresponding ranges of the 2-variable analysis for a 5-year exposure.
On the other hand, the ranges of $\sin^2 \theta_{23}$, both  2$\sigma$ and 3$\sigma$, are the same for both 2 and 3-variable analyses.
When the exposure is increased to 10 years, the parameter ranges are smaller compared to the ranges of 5-year exposure but the relation between the ranges of the 2 and 3-variable analyses are similar to those of 5-year exposure.
We also found that an analysis where the binning is done based on the charge ID of the track also gives the same allowed ranges of oscillation parameters as the analysis without charge ID.

In conclusion, we find that the inclusion of the third variable leads to no noticeable change in the precision of $\sin^2 \theta_{23}$, but improves the precision in $|\Delta m^2_{31}|$ by about 30\% for both 2$\sigma$ and 3$\sigma$ ranges and for both 5 and 10-year exposure times.
The 3-variable analysis method includes particles other than muons produced in the neutrino CC interactions, and hence is more sensitive to the total energy of the event.
The location of the minimum in the L/E$_\nu$ distribution determines $|\Delta m^2_{31}|$.
Hence, a better estimate of the neutrino event energy leads to a better precision in this parameter.
On the other hand, the amount of deficit in the atmospheric $\nu_{\mu}/\bar{\nu}_{\mu}$ flux is a measure of $\sin^2 \theta_{23}$, whose precision is not sensitive to additional kinematic information but does improve with increased  exposure time.

\end{document}